\documentclass[runningheads]{llncs}
\usepackage[T1]{fontenc}
\usepackage{xcolor}   
\usepackage{multirow}
\usepackage{graphicx}
\usepackage{amssymb}
\usepackage{amsmath}
\usepackage[table,xcdraw]{xcolor}
\usepackage{booktabs}
\begin{document}


\title{GeoMER: A Joint Euclidean–Hyperbolic Geometry Framework for Multimodal Emotion Recognition}

\titlerunning{GeoMER}
%

\author{Ting Zhu\inst{1} \and
Huizhi Liang\inst{1} \and
Boguslaw Obara\inst{1} \and
Shufei Duan\inst{2} \and
Varun Ojha\inst{1}}
\authorrunning{T. Zhu et al.}

\institute{Newcastle University, Newcastle upon Tyne, United Kingdom \and
Taiyuan University of Technology, Shanxi, China \\
\email{\{T.zhu11, Huizhi.Liang, Boguslaw.Obara, Varun.Ojha\}@newcastle.ac.uk}\\
\email{duanshufei@tyut.edu.cn}}
%

\maketitle              
\begin{abstract}
Multimodal Emotion Recognition (MER) aims to understand human emotions by integrating information from multiple modalities such as speech and text. Psychological studies suggest that emotions exhibit latent hierarchical structured relationships in their semantic representations~\cite{arano2021multimodal}. However, most MER approaches rely on Euclidean representations, which capture fine-grained cross-modal interactions but do not explicitly provide a geometric prior for organizing hierarchical emotional structures. To mitigate this gap, we propose GeoMER, a joint Euclidean-Hyperbolic modeling framework that introduces hyperbolic emotion-core routing to capture higher-level, abstract emotional representations for MER. The proposed framework jointly models token-level cross-modal semantic interactions and latent emotion representations in two complementary geometric spaces. Specifically, GeoMER consists of a Euclidean Multimodal Fusion Module (EuclFusion) for multimodal representation fusion and a Hyperbolic Geometric Core Routing (HypGeoRoute) module that performs geometry-aware core assignment and aggregation in a hyperbolic manifold to capture more abstract feature representations. The HypGeoRoute module complements the Euclidean token-level modeling. We experiment on the IEMOCAP multimodal MER benchmark and demonstrate that the proposed method achieves the best performance, with 83.54\% weighted accuracy and 83.35\% unweighted accuracy, outperforming Euclidean baselines, including MemoCMT. 

\keywords{Multimodal emotion recognition  \and Hyperbolic geometry \and Speech emotion recognition \and Multimodal feature fusion \and Affective computing.}
\end{abstract}

\section{Introduction}
Multimodal Emotion Recognition (MER) aims to enable machines to interpret human emotions more accurately by integrating multiple modalities, such as speech and text~\cite{wu2025mercsurvey}. Recent advances in deep neural architectures and cross-modal attention mechanisms~\cite{khan2025memocmt,makhmudov2024enhancing} have significantly improved MER performance. However, emotional expressions possess hierarchical structured relationships that go beyond simple categories. Psychological emotion theories~\cite{plutchik1991emotions,russell1980circumplex} suggest that emotions have multi-component hierarchical structures and continuous relationships, supporting emotion mixtures (e.g., joy + anticipation → optimism) and intensity variations (e.g., annoyance $\rightarrow$ anger $\rightarrow$ rage). While modern Transformer-based models can learn expressive representations in Euclidean space, standard Euclidean latent spaces do not explicitly provide a geometric structure for organizing hierarchical emotional structures. In contrast, hyperbolic geometry naturally excels at representing tree-like or multi-level relationships~\cite{nickel2017poincare}, making it a suitable complementary representation space for modeling structured underlying emotional cues.  These limitations motivate exploring non-Euclidean representation spaces to model structured emotion in MER~\cite{chen2023label}.

Hyperbolic geometry has recently attracted attention as a powerful representation space for hierarchical data due to its exponential expansion~\cite{fu2026emotion}. Hyperbolic representation learning has been successfully applied in various domains, including natural language processing and affective computing~\cite{chen2023label,fu2026emotion}. Fine-grained emotion classification~\cite{zhao2025emergence} suggests that emotion categories are not isolated labels, instead exhibiting multi-level and intensity-dependent relationships. HypEmo~\cite{chen2023label} introduces label-aware hyperbolic embeddings to model hierarchical structures among fine-grained emotion categories. Earlier works applied hyperbolic layers to multimodal sentiment and emotion recognition~\cite{arano2021multimodal}, implementing both sequential and parallel mixing schemes between Euclidean and hyperbolic dense layers. Although these studies demonstrate the potential of hyperbolic geometry for affective computing, they focus on the sample's label hierarchy or on general Euclidean and hyperbolic layer composition. On the contrary, our GeoMER is a task-specific dual-geometry design framework.

This design stems from the fact that the MER requires both abstract structure modeling and cross-modal semantic alignment between heterogeneous modalities such as speech and text~\cite{fu2026emotion}. 
At the same time, mapping all representations into hyperbolic space may improve structural modeling, but weaken the advantages of Euclidean representations in cross-modal alignment and introduce an optimization stability problem. Therefore, developing a unified, geometry-aware hybrid model that preserves the strengths of Euclidean representations and hyperbolic geometry is more suitable for MER. Thus, unlike previous works, our work is not a simple replacement of Euclidean embeddings with hyperbolic ones or a direct reuse of these components for MER. Instead, GeoMER is a joint Euclidean-Hyperbolic framework for multimodal emotion recognition. It consists of two branches: a) the Euclidean branch learns discriminative cross-modal representations with cross-attention transformers, and b) the hyperbolic branch models the latent structure of emotions in a hyperbolic manifold. The two branches are jointly optimized and integrated via a unified fusion mechanism, enabling the model to simultaneously preserve both Euclidean discriminative representations and more abstract hyperbolic features. 
The main contributions are as follows:
\begin{enumerate}
    \item We propose GeoMER, a joint Euclidean-Hyperbolic MER framework that specializes both the discriminative semantic modeling (Euclidean space) and abstract hierarchical structural modeling (hyperbolic space).
    \item We propose a HypGeoRoute module that softly assigns token representations to learnable geometric cores in the Poincar\'{e} ball via distance-aware routing.
    \item We propose a core-wise hyperbolic self-attention mechanism that captures relationships among geometric cores in the Poincar\'{e} ball, preserving geometric consistency while enhancing contextual modeling.
\end{enumerate}

\section{Related Work}
Early MER studies~\cite{wu2025mercsurvey} relied on handcrafted features and employed simple fusion methods, which were sensitive to noise. Recent works have used deep neural networks with attention mechanisms~\cite{makhmudov2024enhancing} and Transformer-based architectures such as TDFNet~\cite{zhao2023tdfnet} and MemoCMT~\cite{khan2025memocmt} to capture dynamic interactions across modalities. 

To better capture structured emotional relationships, researchers have introduced hyperbolic representation learning as it can model hierarchical data with exponential capacity~\cite{fu2026emotion}. The hyperbolic Transformer extends the self-attention mechanism to non-Euclidean spaces for hierarchical sequence modeling~\cite{yang2024hypformer}. Several studies have explored the application of hyperbolic modeling in MER, including conversational emotion recognition with hyperbolic hypergraph fusion~\cite{zheng2025multimodal}, hyperbolic fusion of heterogeneous speech representations~\cite{phukan2025hyfuse}, and label-aware hyperbolic embeddings for fine-grained emotion classification~\cite{chen2023label}. Meanwhile, hybrid geometric learning has been preliminarily studied, in which Euclidean encoders are combined with hyperbolic components for multimodal representation learning~\cite{arano2021multimodal}. Emotion Collider~\cite{fu2026emotion}, HypEmo~\cite{chen2023label}, and HyperPoint~\cite{sun2025hyperpoint} further explore hyperbolic modeling for emotion mixture, label-aware classification, and multimodal representation learning. 

However, there are two key limitations. First, most existing MER methods operate in Euclidean space, which struggles to capture the more structured and abstract emotional representations. Although hyperbolic approaches have been explored, they remain limited in scope and often focus on specific components, such as embeddings or labels. Second, hybrid geometric methods remain underdeveloped, typically combining Euclidean and hyperbolic representations in a loosely coupled manner without a unified framework for joint optimization.

\section{Methodology}
\begin{figure*}[t]
\centering
\includegraphics[width=\textwidth]{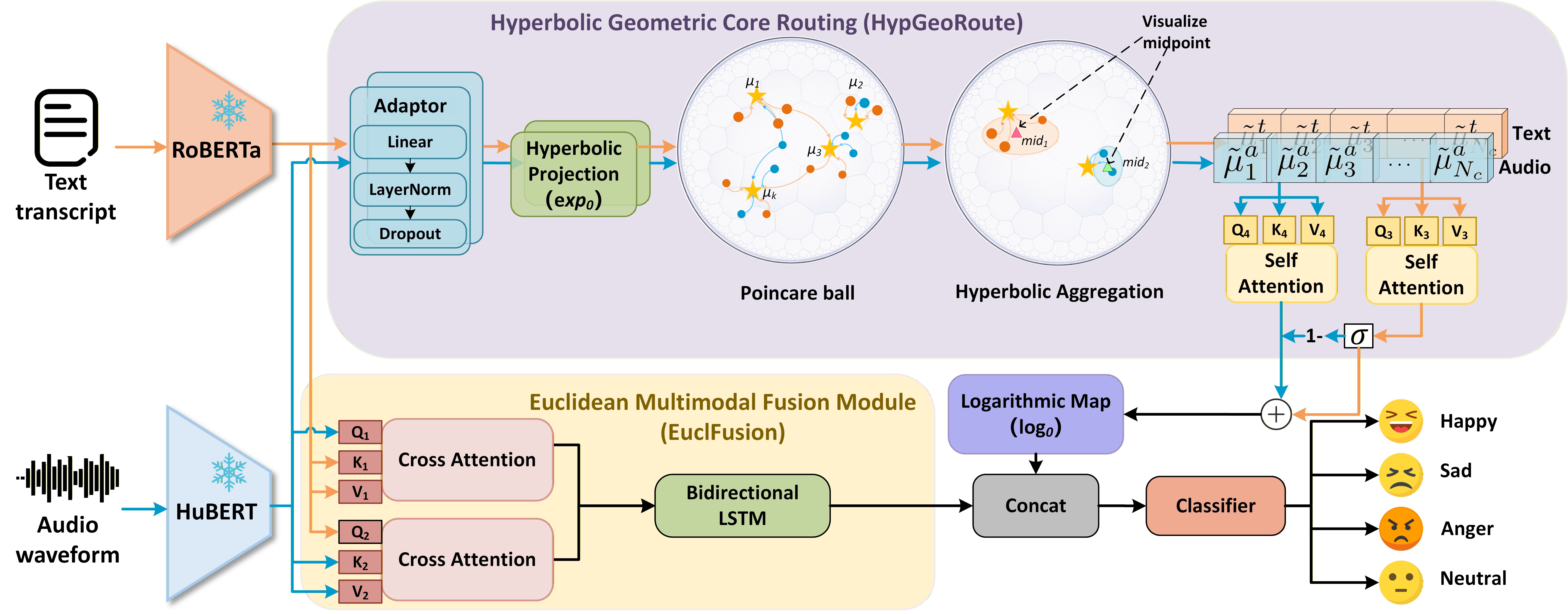}
\caption{Overview of the GeoMER framework. Text and audio features extracted by pre-trained encoders are first aligned by a Euclidean adaptor. EuclFusion models multimodal interactions in Euclidean space, while HypGeoRoute captures abstract emotional features via geometric core routing in a hyperbolic manifold. The resulting hyperbolic representations are mapped back to Euclidean space and concatenated with Euclidean features for the final emotion
classification.}
\label{fig:framework}
\end{figure*}
We propose GeoMER, a Euclidean–Hyperbolic modeling framework for MER that models discriminative cross-modal semantics in Euclidean space and structured emotion representations on a hyperbolic manifold. As shown in Figure~\ref {fig:framework}, the framework consists of feature extraction, Euclidean multimodal fusion, hyperbolic core routing, and final emotion classification. Specifically, \emph{Euclidean Multimodal Fusion Module (EuclFusion)} models cross-modal interactions, and the \emph{Hyperbolic Geometric Core Routing (HypGeoRoute)} module projects multimodal token representations onto a hyperbolic manifold and organizes them around learnable geometric cores via geometry-aware routing and core-wise aggregation. The final representation is obtained by concatenating the two representations, enabling GeoMER to jointly exploit multimodal interactions and abstract geometric features for emotion prediction.

\subsection{Feature Embedding}
We use the HuBERT encoder to encode the raw audio into a series of acoustic embeddings. All audio signals are 16 kHz. The RoBERTa model is employed to obtain contextual tokens from the transcriptions.

\subsection{Euclidean Multimodal Fusion Module}
The EuclFusion is designed to model cross-modal interactions in Euclidean space. The architecture follows MemoCMT~\cite{khan2025memocmt} while improving on simple concatenation by introducing a bidirectional Long Short-Term Memory (BiLSTM) network for sequential aggregation. Two multi-head cross-attention mechanisms are first employed to effectively fuse speech and text cues. This step enhances token-level alignment across modalities. While cross-attention focuses on aligning and exchanging information between speech and text tokens, the subsequent BiLSTM serves a different role by modeling the temporal evolution of the fused multimodal sequence at the utterance level. Specifically, the fused multimodal token sequence is processed by a BiLSTM. This step dynamically integrates cross-modal cues over time, which is crucial for capturing the evolving nature of emotional expression in speech~\cite{lee2020multimodal,li2021learning}.  Let the audio and text token sequences $V_a \in \mathbb{R}^{n_a \times d}$ and $V_t \in \mathbb{R}^{n_t \times d}$, where $a$ is for audio and $t$ is for text, $n_a$ and $n_t$ are the token lengths, and $d$ is the embedding dimension. The resulting hidden states are aggregated to produce a fixed-length utterance-level embedding $h_{euc} \in \mathbb{R}^{d}$. This Euclidean representation captures multimodal interactions in embedding space and provides complementary information to the subsequent hyperbolic modeling branch.

\subsection{Hyperbolic Geometric Core Routing}
The HypGeoRoute module is designed to capture latent abstract information through geometry-aware token routing in a hyperbolic manifold. Unlike traditional Euclidean representation modeling, HypGeoRoute operates in the Poincar\'{e} ball to organize multimodal token embeddings around learnable geometric cores. Tokens are softly routed to multiple cores based on hyperbolic distances, enabling geometrically consistent aggregation of multimodal abstract information.

\subsubsection{Feature Alignment and Hyperbolic Emotion Representation.}
The audio and text token embeddings $V_a \in \mathbb{R}^{n_a \times d}$ and $V_t \in \mathbb{R}^{n_t \times d}$ extracted from frozen pre-trained encoders have modality-specific scale and distribution mismatches. To stabilize subsequent geometric operations, we introduce lightweight Euclidean adaptors composed of a linear projection (Linear), layer normalization (LN), and dropout (Drop). Let $W_a, W_t \in \mathbb{R}^{d \times d}$ and $b_a, b_t \in \mathbb{R}^{d}$ denote learnable projection parameters for audio and text modalities. The adapted audio and text embeddings $\tilde{V}_a$ and $\tilde{V}_t$ are therefore computed as:
\begin{equation}
\tilde{V_a} = \mathrm{Drop}\big(\mathrm{LN}(V_a W_a + b_a)\big), \quad
\tilde{V_t} = \mathrm{Drop}\big(\mathrm{LN}(V_t W_t + b_t)\big).
\end{equation}
These audio and text features are then projected onto the Poincar\'{e} ball model $\mathbb{B}_c^d$ via an exponential map at the origin as: $ H_a = \exp_0(\tilde{V_a})$ and $ H_t = \exp_0(\tilde{V_t})$, where 
\begin{equation}
\exp_0(\mathbf{x}) =
\tanh\!\left(\sqrt{c}\,\|\mathbf{x}\|\right)
\frac{\mathbf{x}}{\sqrt{c}\,\|\mathbf{x}\|},
\end{equation}
$\mathbf{x}\in\mathbb{R}^d$ is $\tilde{V_a}$ for audio, and  $\mathbf{x}\in\mathbb{R}^d$ is  $\tilde{V_t}$ for text. We use $c>0$ to control the magnitude of the negative curvature.

\subsubsection{Mixed Abstract Emotion Representation via Geometric Core Routing.}
Traditional Euclidean approaches typically aggregate all tokens into a single utterance-level vector with a single emotion label. However, in real-world conversations, it may contain multiple emotional cues with varying degrees of dominance, resulting in an emotional mixture rather than a single, discrete emotion. Therefore, we introduce $N_c$ learnable geometric cores $\{\mu_k \mid k = 1,\ldots,N_c\}  \subset \mathbb{B}_c^d$, shared across modalities and acting as latent emotional components in hyperbolic space. This routing mechanism organizes token-level emotional cues into higher-level emotional components, enabling the model to learn abstract emotional representations in a hyperbolic manifold. For each modality $m \in \{a,t\}$ with hyperbolic token embeddings $H_m = \{h_i^{m}\}_{i=1}^{L_m}$, where $L_m$ denotes the sequence length of modality $m$, we compute the hyperbolic distance $d_{ik}^{m}$ between a token $i$ and a core $k$ in the Poincar\'{e} ball as:
{\small
\begin{equation}
d_{ik}^{m}=\mathrm{arcosh} \left(1+ \frac{
2\left\|h_i^{m}-\mu_k\right\|^2
}{
\left(1-\left\|h_i^{m}\right\|^2\right)
\left(1-\left\|\mu_k\right\|^2\right)
}
\right),
\label{eq:hyper_dist}
\end{equation}}
where $\|\cdot\|$ denotes the Euclidean norm in the ambient space.

Instead of assigning each utterance to a single core, tokens are softly routed to multiple cores according to their hyperbolic distance, allowing multiple tokens (utterances) to form a hierarchical abstract representation in hyperbolic space. The routing weight $\alpha_{ik}^{m}$ of token $i$ to core $k$ under modality $m$ is defined as:
\begin{equation}
\alpha_{ik}^{m} = \frac{
\exp\!\left(-d_{ik}^{m}\right)
}{
\sum_{j=1}^{N_c} \exp\!\left(-d_{ij}^{m}\right) 
},
\quad
\sum_{k=1}^{N_c}
\alpha_{ik}^{m} = 1.
\label{eq:routing}
\end{equation}
where $j$ iterates over all $N_c$ cores for normalization.

\subsubsection{Token Aggregation for Hyperbolic Cores.}
Given the routing weights $\alpha_{ik}^{m}$, we aggregate modality-specific token embeddings into $N_c$ core-level representations. For each modality $m$ and core $k$, we employ the Einstein midpoint~\cite{ganea2018hyperbolic} for computing a geometrically consistent weighted mean in the 
Poincar\'{e} ball. Specifically, hyperbolic token embeddings are first mapped to the tangent space at the origin using $\log_{\mathbf{0}}(\cdot)$. The routing-weighted aggregation is then performed in this space, followed by projection back to the hyperbolic manifold via $\exp_{\mathbf{0}}(\cdot)$:
\begin{equation}
\tilde{\mu}_k^{m}
= \exp_{\mathbf{0}}\!\left(
    \sum_{i=1}^{L_m}
    \alpha_{ik}^{m}\,
    \log_{\mathbf{0}}\!\left({h}_i^{m}\right)
\right),
\label{eq:einstein_midpoint}
\end{equation}
where $h_i^{m} \in \mathbb{B}_c^d$ is the 
$i$-th hyperbolic token embedding of modality $m$, 
and $\tilde{\mu}_k^{m} \in \mathbb{B}_c^d$ is the aggregated core representation for the $k$-th abstract emotion prototype under modality $m$.
This operation is applied for each modality $m \in \{a, t\}$ and each core $k \in \{1, \ldots, N_c\}$, obtaining two sets of core-level representations: 
\begin{equation}
M^m = 
\left[\,
\tilde{\mu}_1^{m},\,
\tilde{\mu}_2^{m},\,
\tilde{\mu}_3^{m},\,
\ldots,\,
\tilde{\mu}_{N_c}^{m}
\,\right].
\label{eq:core_matrix}
\end{equation}

\subsubsection{Hyperbolic Self-Attention over Emotion Cores.}
After getting the core-level representations $M^m \in \mathbb{B}_c^{N_c \times d}$ for each modality $m \in \{a, t\}$, we apply Hyperbolic Self-Attention (HSA) across the $N_c$ emotion cores to model inter-core semantic correlations between different abstract emotion prototypes. This mechanism enables interactions among latent emotional features, allowing related abstract emotional features to influence one another on the hyperbolic manifold. Since linear transformations are not directly defined on Riemannian manifolds, Queries (Q), Keys (K), and Values (V) are projected by Poincar\'{e} linear layers ($\log_{\mathbf{0}} \to $ Linear $\to \exp_{\mathbf{0}}$), where $W_Q, W_K, W_V \in \mathbb{R}^{d \times d}$ are learnable projection matrices:
\begin{equation}
\begin{aligned}
Q^{m} &= \exp_{\mathbf{0}}\!\left(
    \log_{\mathbf{0}}\!\left(M^m\right)W_Q 
\right), \quad
K^{m} = \exp_{\mathbf{0}}\!\left(
    \log_{\mathbf{0}}\!\left(M^m\right)W_K
\right),  and \\
V^{m} &= \exp_{\mathbf{0}}\!\left(
    \log_{\mathbf{0}}\!\left(M^m\right)W_V
\right).
\end{aligned}
\label{eq:qkv_proj}
\end{equation}

We adopt a multi-head mechanism with $R$ attention heads. For the $r$-th head, the attention score $s_{pq}^{(m,r)}$ and attention weights $a_{pq}^{(m,r)}$ (for $r=1, 2,..., R$) between the $p$-th query core and the $q$-th key core  are defined as: 
\begin{equation}
s_{pq}^{(m,r)} = 
-\frac{d_{\mathbb{B}}\!\left(
Q_{p}^{(m,r)},\, K_{q}^{(m,r)}
\right)}{\exp(\tau)},
\quad
a_{pq}^{(m,r)} =
\frac{\exp(s_{pq}^{(m,r)})}
{\sum_{q'=1}^{N_c}\exp(s_{pq'}^{(m,r)})},
\label{eq:hsa_score}
\end{equation}
where $d_{\mathbb{B}}(\cdot,\cdot)$ denotes the geodesic distance on the Poincar\'{e} ball, and $\tau$ is a learnable parameter, called temperature in literature.

Given the value representations $V_q^{(m,r)}$, the output $O_{p}^{(m,r)}$ is computed using the Einstein midpoint as:
\begin{equation}
O_{p}^{(m,r)} =
\mathrm{HypAgg}\!\left(
\{V_q^{(m,r)}\}_{q=1}^{N_c},
\{a_{pq}^{(m,r)}\}_{q=1}^{N_c}
\right),
\label{eq:hsa_agg}
\end{equation}
where $\mathrm{HypAgg}(\cdot)$ denotes weighted midpoint aggregation on the manifold. 

The outputs of all attention heads are first concatenated and projected through a hyperbolic linear transformation $\mathrm{HypLinear(\cdot)}$, and then the resulting representation is combined with the input $M^m$ through a hyperbolic residual connection using M\"obius addition. The output $Z_m$ of HSA is defined as:
\begin{equation}
Z_m =
\frac{1}{\sqrt{2}}\,\otimes
\left(M^m \oplus
\mathrm{HypLinear}\!\left(
\mathrm{Concat}(O^{(m,1)},\ldots,O^{(m,R)})
\right)
\right),
\label{eq:hsa_out}
\end{equation}
where $\oplus$ denotes M\"obius addition and $\otimes$ denotes M\"obius scalar multiplication. This residual design preserves manifold consistency under M\"obius operations while stabilizing feature propagation across layers.

\subsubsection{Cross-modal Gated Fusion.}
After core-wise hyperbolic self-attention, each modality produces $N_c$ core-specific abstract emotion representations. Given $Z_{t} = \{z_k^{t}\}_{k=1}^{N_c}$ and $Z_{a} = \{z_k^{a}\}_{k=1}^{N_c}$, we fuse the text and audio representations of each core $k$ directly in the Poincar\'{e} ball through M\"obius operations, ensuring that the representations remain on the hyperbolic manifold. For each core $k$, we compute a gate signal $g_k\in (0,1)$ from the difference of two modality representations in the tangent space at the origin, which measures cross-modal inconsistency:
\begin{equation}
g_k =
\sigma\!\left(W_g
\left(|\log_{\mathbf{0}}\!\left(z_k^{t}\right)
-
\log_{\mathbf{0}}\!\left(z_k^{a}\right)|
\right)
+ b_g
\right),
\label{eq:gate}
\end{equation}
where $W_g \in \mathbb{R}^{d \times d}$ and $b_g \in \mathbb{R}^d$ are learnable parameters, and $\sigma(\cdot)$ denotes the sigmoid function. A larger discrepancy produces a stronger gating response, allowing the model to emphasize the more informative modality. 

The fused representation of core $k$ is then computed using M\"obius scalar multiplication and M\"obius addition:
\begin{equation}
f_k =
\frac{1}{\sqrt{2}}
\otimes
\left(
(g_k \otimes z_k^{t})
\oplus
((1-g_k)\otimes z_k^{a})
\right),
\label{eq:mobius_fusion}
\end{equation}
where $\otimes$ and $\oplus$ denote M\"obius scalar multiplication and M\"obius addition, respectively. 

The fused core representations $F^{hyp} = \{f_k\}_{k=1}^{N_c}$ are mapped back into Euclidean space to get $F^{euc}= \{f_k^{euc}\}_{k=1}^{N_c}$ via the logarithmic map function. Since an utterance may express multiple latent emotional cues with different contributions, we further estimate the importance of each core at the utterance level.  Therefore, we compute a core importance distribution $\{\pi_k\}_{k=1}^{N_c}$  using the softmax function, which reflects how strongly each emotional core contributes to the overall emotional representation. The final representation of the hyperbolic branch ($h_{hyp}$) is obtained by weighted aggregation of the fused cores as:
\begin{equation}
h_{hyp} =
\sum_{k=1}^{N_c}
\pi_k\,
f_k^{{euc}},
\label{eq:core_agg}
\end{equation}
where $\pi_k$ denotes the importance weight of the $k$-th core.

Finally, we concatenate the representations from the Euclidean branch ($h_{euc}$) and the hyperbolic branch ($h_{hyp}$) and then feed them into a linear classifier followed by a softmax function:
\begin{equation}
\hat{y} =\mathrm{Softmax}
\big(
W_c \,
\mathrm{Concat}(h_{euc}, h_{hyp}) + b_c
\big),
\end{equation}
where $Concat(\cdot)$ is the concatenation operation, $W_c$ and $b_c$ are learnable classifier parameters, and $\hat{y}$ is the predicted probability distribution over emotion classes.
\section{Experiments}
\subsubsection{MER Benchmark}
We use the IEMOCAP benchmark~\cite{busso2008iemocap} that consists of 12 hours of audio-visual emotional conversations performed by 10 professional actors (5 male, 5 female). Each session includes both scripted and improvised dialogues designed to elicit natural emotional expressions. It provides multimodal recordings, including audio, text, facial motion capture, and videos. Only speech and text modalities are used in this paper. Following prior studies, four emotion classes, including \textit{angry}, \textit{happy}, \textit{sad}, and \textit{neutral}, are chosen. IEMOCAP contains 5,531 utterances in total, 4479 training set, 498 validation samples, and 554 test samples. The distribution of the training set, test set, and validation set is shown in Table~\ref{tab:iemocap_distribution}. It can be observed that the IEMOCAP is moderately class-imbalanced,  where the \textit{happy} and \textit{neutral} categories contain more samples than \textit{angry} and \textit{sad}.
\begin{table}[t]
\centering
\caption{Emotion distribution of IEMOCAP under the four-class setting. Percentages are computed within each split.}
\label{tab:iemocap_distribution}
\begin{tabular}{lcccccccc}
\hline
\multirow{2}{*}{Emotion} & \multicolumn{2}{c}{Train} & \multicolumn{2}{c}{Val} & \multicolumn{2}{c}{Test} & \multicolumn{2}{c}{Total} \\
\cline{2-9}
& \#Utterances & (\%) & \#Utterances & (\%) & \#Utterances & (\%) & \#Utterances & (\%) \\
\hline
Angry & 902 & 20.1 & 99 & 19.9 & 102 & 18.4 & 1103 & 19.9 \\
Happy & 1321 & 29.5 & 144 & 28.9 & 171 & 30.9 & 1636 & 29.6 \\
Sad & 857 & 19.1 & 98 & 19.7 & 129 & 23.3 & 1084 & 19.6 \\
Neutral & 1399 & 31.2 & 157 & 31.5 & 152 & 27.4 & 1708 & 30.9 \\
\hline
Total & 4479 & 100 & 498 & 100 & 554 & 100 & 5531 & 100 \\
\hline
\end{tabular}
\end{table}

\subsubsection{Evaluation Metrics.}
Given the class imbalance in IEMOCAP, we evaluate model performance using both overall and class-balanced metrics. Weighted accuracy (WA), unweighted accuracy (UA), and Macro-F1 are reported. WA measures the overall classification accuracy, while UA averages recall across all classes to better reflect performance under class imbalance. In addition, Macro-F1 is obtained by averaging the F1-scores of all classes.

\subsubsection{Experiment Setup.}
All experiments are implemented in PyTorch and conducted on one NVIDIA RTX 3090 GPU. HuBERT-base and RoBERTa-base are used as frozen acoustic and text encoders, producing 768-dimensional features. The model is optimized using AdamW with $\beta_1=0.9$, $\beta_2=0.999$, $\epsilon=10^{-8}$, and weight decay $5\times10^{-3}$, and the cross-entropy loss. The initial learning rate is $1\times10^{-4}$ and decayed by a factor of 0.1 every 30 epochs. The model is trained for up to 60 epochs with early stopping based on validation UA. The hyperbolic curvature is $c=1.0$, the hyperbolic routing core is $N_c=4$, and the scaling coefficient $\alpha$ is learnable. 

\subsection{Main Results}
\begin{table}[t]
\caption{Comparison with representative multimodal methods on IEMOCAP (4-class). Results marked with $^\dagger$ are reproduced under our experimental setting and reported as mean $\pm$ standard deviation over five runs; other results are taken from the corresponding published papers. WA and UA indicated the weighted and unweighted accuracies.}
\label{tab:iemocap_sota}
\centering
\begin{tabular}{lcc}
\hline
Model  & WA (\%) & UA (\%) \\
\hline
\rowcolor{gray!12}
\multicolumn{3}{l}{\textbf{Audio-only Modality}}\\
CNN  & 65.30 & 64.20 \\
LSTM  & 66.80 & 65.90 \\
MemoCMT (Audio-only)~\qquad\quad &  66.56 & 65.34\\
GeoMER (Audio-only) &  70.80 & 70.36 \\
\rowcolor{gray!12}
\multicolumn{3}{l}{\textbf{Text-only Modality}} \\
BERT & 70.50 & 69.30 \\
MemoCMT (Text-only) &  71.10 & 71.12\\
GeoMER (Text-only) &  73.24 & 72.89 \\
\rowcolor{gray!12}
\multicolumn{3}{l}{\textbf{Text + Audio Modalities}} \\
Fan~\cite{Fan2023MGAT} & 78.51 & 79.31 \\
Sun~\cite{sun2023using} & 78.42 & 79.71 \\
Lei~\cite{lei2026scmi} & 78.60 & 77.60 \\
MemoCMT(reported)~\cite{khan2025memocmt} & 81.85 & 81.33 \\
MemoCMT(reproduced)$^\dagger$ & 81.05$\pm$0.45 & 81.33$\pm$0.34 \\
\textbf{HypGeoRoute(Ours)} & \textit{82.78$\pm$0.71} & \textit{82.64$\pm$0.67} \\
\textbf{GeoMER(Ours)} & \textbf{83.54$\pm$0.77} & \textbf{83.35$\pm$0.57} \\
\hline
\end{tabular}
\end{table}
Table~\ref{tab:iemocap_sota} presents the comparison between the proposed method and several representative baselines on the IEMOCAP benchmark under the four-class setting. When only acoustic features are used, our GeoMER achieves 70.80\% WA and 70.36\% UA, outperforming classical audio baselines such as CNN and LSTM. Compared with MemoCMT (Audio-only), our method improves WA by 4.24\% and UA by 5.02\%, indicating its benefit even in a single-modality setting. For text-based emotion recognition, GeoMER achieves 73.24\% WA and 72.89\% UA, outperforming both BERT and MemoCMT (Text-only). This demonstrates that it effectively captures semantic emotional cues from textual inputs. When using the hyperbolic-only (HypGeoRoute), the model achieves competitive performance, with 82.78\% WA and 82.64\% UA, outperforming the Euclidean baseline, MemoCMT. When both audio and text modalities are used, the proposed GeoMER achieves the best performance, with 83.54 $\pm$ 0.77\% WA and 83.35 $\pm$ 0.57\% UA across five runs. Compared with the reported Euclidean baseline MemoCMT result, our method improves WA by 1.69\% and UA by 2.02\%. Compared with our reproduced MemoCMT baseline, GeoMER improves WA by 2.49\% and UA by 2.02\%. These results demonstrate the effectiveness of multimodal fusion and evidence that incorporating hyperbolic geometry benefits multimodal representation learning. Our method models abstract, latent relationships among emotional representations via hyperbolic geometric routing. Therefore, GeoMER does not replace Euclidean representation learning; instead, it complements it by modeling hyperbolic structure. Overall, the results verify the advantage of this hybrid Euclidean–hyperbolic design for multimodal emotion representations.

\subsection{Ablation Study}

Table~\ref{tab:ablation} presents the ablation results of the proposed components on the IEMOCAP benchmark. Based on the MemoCMT baseline (81.05\% WA and 81.33\% UA), adding the Euclidean Multimodal Fusion Module (EuclFusion) improves the performance to 81.95\% WA and 82.38\% UA. This corresponds to an improvement of 0.90\% in WA and 1.05\% in UA, demonstrating that enhanced cross-modal interaction can effectively improve emotion representation learning. We further explore several variants of hyperbolic modeling. When the Hyperbolic Geometric Core Routing (HypGeoRoute) module is further incorporated, the full GeoMER model achieves the best performance with 83.54\% WA and 83.35\% UA. Compared with the MemoCMT baseline, the proposed method improves WA by 2.49\% and UA by 2.02\%.  Similar improvements are also observed on Macro-F1, which increases from 80.84\% to 83.17\%. These results indicate that the hyperbolic routing mechanism can effectively capture latent hierarchical structures in emotional representations. All ablation results show that Euclidean multimodal fusion and hyperbolic geometric modeling play complementary roles.  The Euclidean branch strengthens speech and text cross-modal interaction, and the hyperbolic branch provides a geometry-aware structure for organizing latent abstract representations.  This supports the effectiveness of the hybrid Euclidean--hyperbolic design for multimodal emotion recognition.
\begin{table}[t]
\centering
\caption{Ablation study of different components on IEMOCAP (4-class). WA, UA, and Macro-F1 denote weighted accuracy, unweighted accuracy, and macro-averaged F1-score, respectively.  $\Delta$WA/$\Delta$UA reports the absolute gain over the reproduced MemoCMT baseline.}
\label{tab:ablation}
\begin{tabular}{lccccc}
\hline
Model & WA (\%) & UA (\%) & Macro-F1 (\%) & $\Delta$WA/$\Delta$UA \\
\hline
MemoCMT(reproduced) & 81.05 & 81.33 & 80.84  & -- \\
+EuclFusion & 81.95 & 82.38 & 81.77 & +0.90/+1.05 \\
+Hyperbolic mean pooling  & 82.16 & 82.53 & 81.95 & +1.11/+1.20 \\
+Hyp self-attention (w/o Routing) & 82.51 & 82.99 & 82.31 & +1.46/+1.66 \\
+Hyp cross-attention (w/o Routing) & 82.54 & 82.60 & 82.42 & +1.49/+1.27 \\
\textbf{GeoMER (+HypGeoRoute)} \textbf{(Ours)} & \textbf{83.54} & \textbf{83.35} & \textbf{83.17} & +2.49/+2.02 \\
\hline
\end{tabular}
\end{table}

\begin{figure}[t]
\centering
\includegraphics[width=\textwidth]
{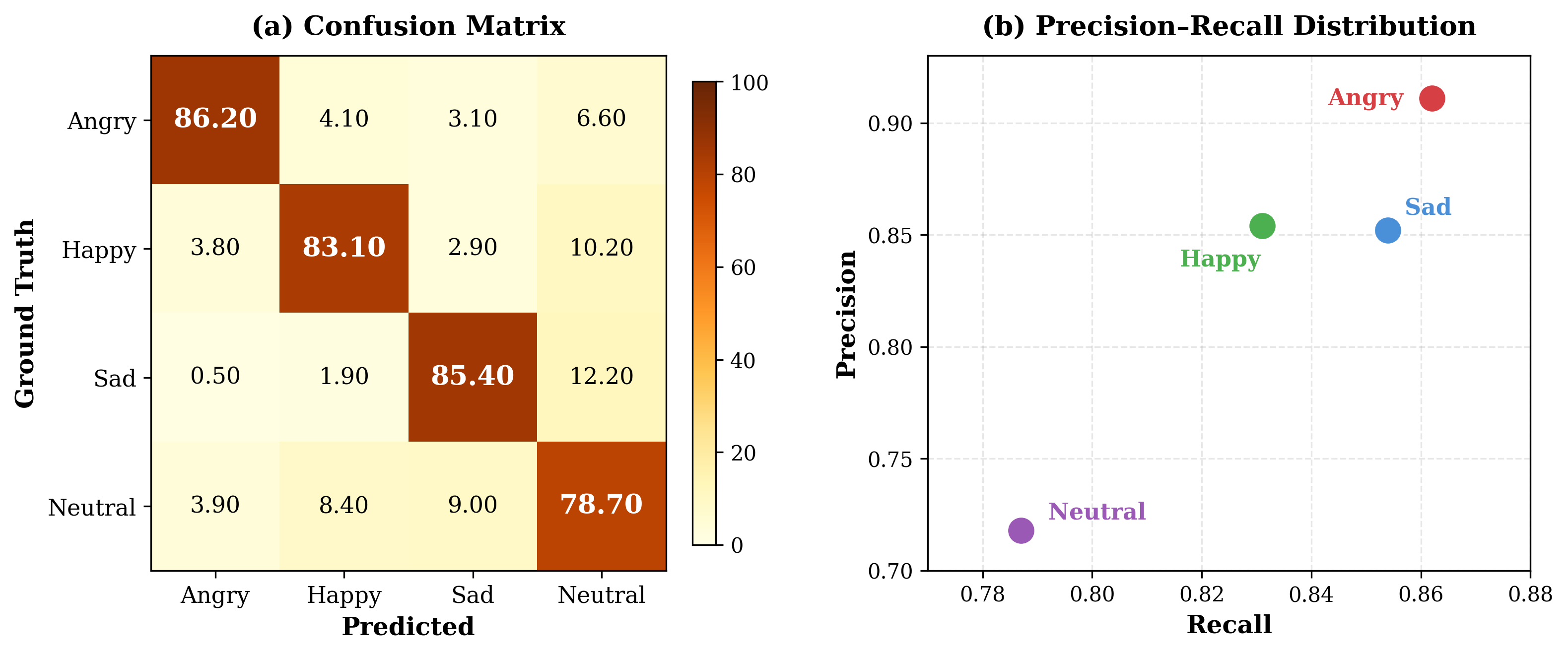}
\caption{Per-class performance analysis of GeoMER on the IEMOCAP benchmark. (a) Confusion matrix, where rows denote ground-truth labels and columns denote predicted labels. The values indicate the classification percentage for each emotion category. (b) Precision--recall distribution of each emotion category.}
\label{fig:per_class}
\end{figure}
Figure~\ref{fig:per_class}~(a) shows the confusion matrix of GeoMER on the IEMOCAP benchmark. Angry achieves the highest recall of 86.20\%, followed by Sad (85.40\%) and Happy (83.10\%). Neutral remains the most challenging category, with a recall of 78.70\%. Specifically, neutral samples are more frequently misclassified as Sad (9.00\%) and Happy (8.40\%), reflecting the inherent ambiguity of neutral emotional expressions in natural conversations.  As shown in Figure~\ref{fig:per_class}~(b), the precision–recall distribution for each emotion category on the IEMOCAP benchmark has the same pattern. Angry achieves the best overall balance, with 91.10\% precision and 86.20\% recall. Sad also shows strong performance with 85.20\% precision, and Happy reaches 85.40\% precision. While Neutral obtains the lowest precision (71.80\%), mainly because samples from other categories are often predicted as Neutral. Overall, GeoMER achieves relatively balanced recognition across all four categories, with a per-class recall range of only 7.50\%. This suggests that the proposed hyperbolic core routing mechanism helps the model capture more discriminative representations.

\begin{figure}[t]
\centering
\small
\includegraphics[width=0.8\textwidth]{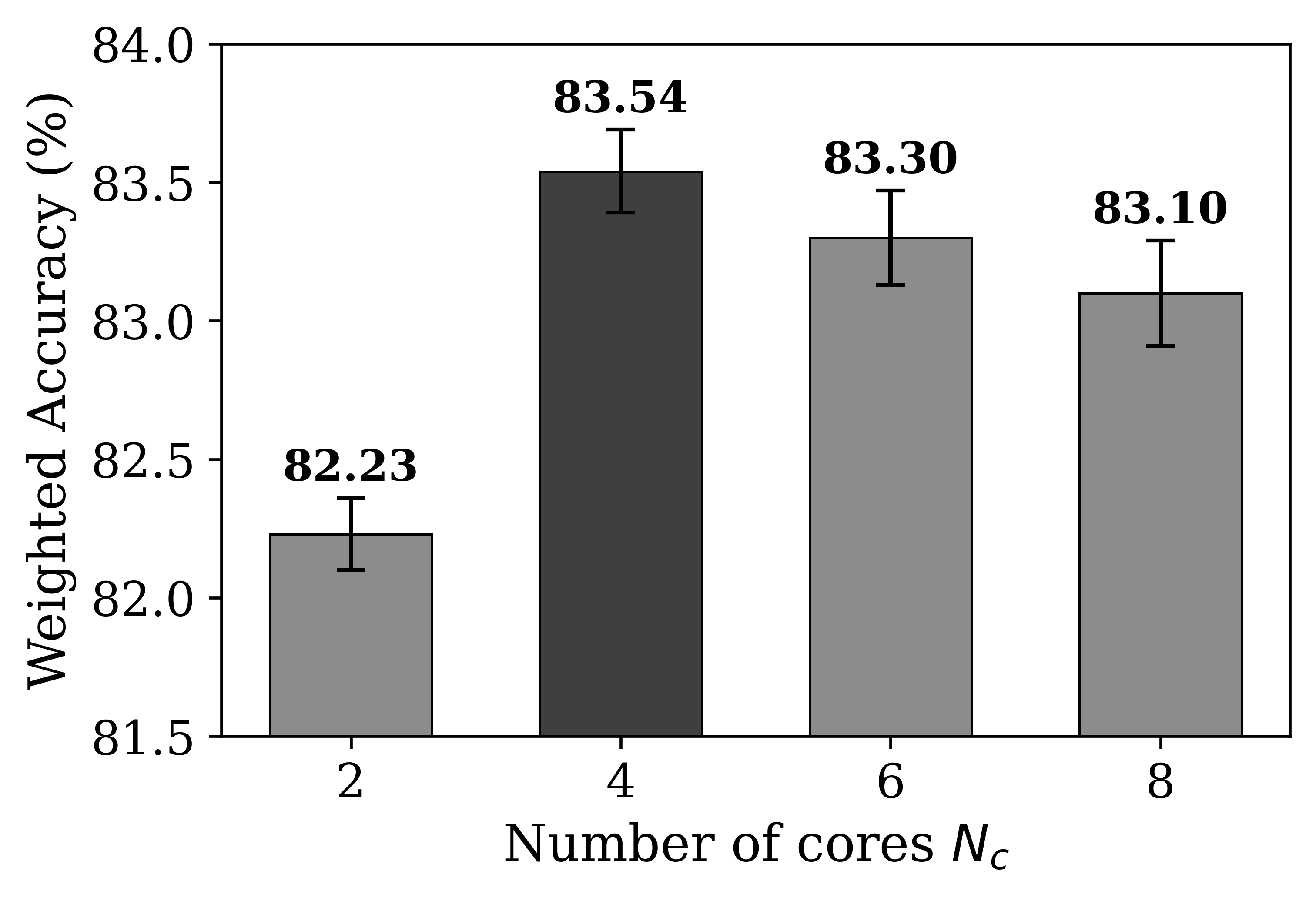}
\caption{Effect of the number of emotion cores $N_c$ in HypGeoRoute on weighted accuracy(\%). Performance is best when $N_c=4$, but larger values also yield good results. This suggests that the learned emotion cores are not restricted to a strict one-to-one mapping to emotion classes, but instead serve as latent affective anchors for flexible representation learning.}
\label{fig:core_analysis}
\end{figure}
As shown in Figure~\ref{fig:core_analysis}, we further analyze the effect of the number of emotion cores $N_c$ in HypGeoRoute. Performance improves WA from 82.23\% to 83.54\% as $N_c$ increases from 2 to 4, suggesting that a small number of cores may be insufficient to capture diverse affective structures. Larger values ($N_c = 6$ or $8$) are slightly lower but still achieve competitive performance, with weighted accuracies of 83.30\% and 83.10\%, respectively. It indicates that the performance of GeoMER does not strictly rely on a one-to-one correspondence between the number of cores and the number of emotion classes. Instead, the learned cores serve as latent anchors that support soft routing and flexible representation learning. In addition, the soft routing mechanism assigns each token to all cores with continuous weights via softmax, so tokens from a single utterance naturally distribute across multiple cores rather than to one. Therefore, we set $N_c = 4$ in the final model, as it achieves the best balance between performance and model complexity.

\section{Conclusion}

In this paper, we propose GeoMER, a dual Euclidean–hyperbolic framework for multimodal emotion recognition. The proposed method combines a Euclidean Multimodal Fusion Module (EuclFusion) for modeling cross-modal interactions with a Hyperbolic Geometric Core Routing (HypGeoRoute) module to capture latent hierarchical structures in emotional representations. Unlike existing hyperbolic emotion embedding methods that mainly focus on label-level or representation-level modeling, distance-aware assignment, manifold-consistent aggregation, core-wise hyperbolic self-attention, and cross-modal gated fusion on the Poincar\'{e} ball, our HypGeoRoute module improves performance by capturing hierarchical structure across modalities. By integrating Euclidean multimodal fusion with hyperbolic geometric modeling, GeoMER provides a more expressive representation for complex emotional structure. Experimental results on the IEMOCAP benchmark show that the proposed method performs well, achieving 83.54\% WA and 83.35\% UA, outperforming the MemoCMT baseline by 2.49\% WA and 2.02\% UA. The ablation study further validates the effectiveness of the proposed components. 


%
%
%
\bibliographystyle{splncs04}
\bibliography{mybibliography}

\end{document}